\documentstyle[11pt,paspconf,epsf]{article}

\begin{document}

\title{Cosmic Star Formation from the Milky Way and its 
Satellites}

\author{F.D.A. Hartwick}
\affil{Department of Physics and Astronomy, University of Victoria,
Victoria, BC, Canada, V8W 3P6}

\begin{abstract}
We use observations and evolutionary models of local objects
to interpret a recent determination of the star-formation history 
of the universe. By fitting the global star-formation rate, 
the model predicts the ratio of spheroid to disk
mass of $\sim$1, an intergalactic medium (IGM) whose mass is $\sim$2.3 times the mass in stars, and
whose metallicity is $\sim$0.1 Z$_\odot$. 
\end{abstract}


\section{Introduction}

It is conventional to interpret observations of high-redshift galaxies 
using sophisticated galaxy evolution models. However, unless the observed
spectra are of sufficiently high S/N, the resulting interpretations may
not be unique (c.f. O'Connell 1996). On the other hand, deep color-magnitude diagrams of objects like globular clusters, dwarf spheroidals,
and gas-rich systems such as the Galactic disk and the Magellanic Clouds
in principle show us their complete star formation history (c.f. 
contributions in Leitherer et al.\ 1996 and this volume). To the extent 
that these systems can be considered representative of the universe as a
whole (the mean density of the local group is close to that of the universe), we should be able to use them to deduce a global star-formation history.
 
\section{The Cosmic Star-Formation History}

\begin{figure}
\plotfiddle{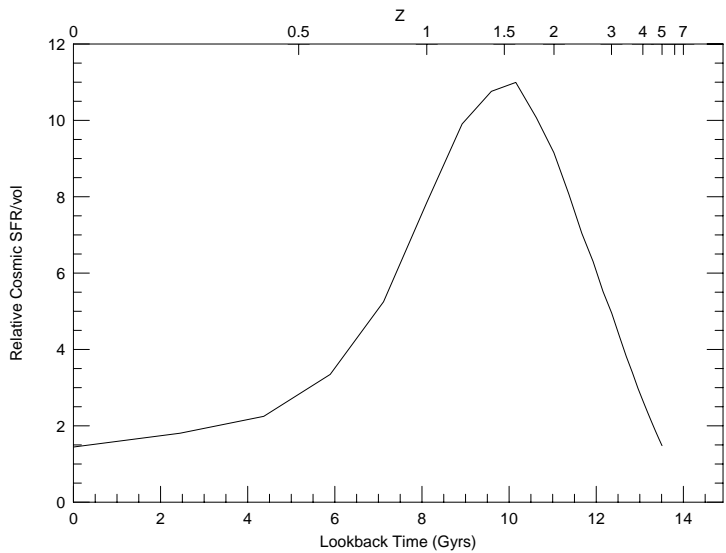}{2.5in}{90}{40}{40}{144}{-25}
\caption{The relative star-formation rate density of Madau et al.\ (1998) plotted as a function of lookback time.}
\end{figure}

The starting point for this work is the diagram of Madau et al.\ (1996) 
and an up-dated version of this diagram from Madau et al.\ (1998) which
shows the cosmic star-formation rate (SFR) in units of M$_\odot~$yr$^{-1}$vol$^{-1}$ as  
a function of redshift z. Very similar results had been derived previously 
by Pei \& Fall (1995) by modelling the observed column densities of H~I and
their metallicities in damped Ly$\alpha$ systems. Because we will use the
results of stellar evolution calculations as input, it is convenient to 
transform the abscissa from z to lookback time by assuming a specific
cosmological model. (The ordinate must also be modified slightly to take 
into account the model change.) The model assumed has $H_0$$=$70 km sec$^{-1}$Mpc$^{-1}$, $\Omega_0$$=$0.2, $\Omega_\Lambda$$=$0.8, and $t_0$$=$14.91 Gyr. This transformed cosmic SFR with a rescaled ordinate is shown in Fig. 1.

\section{Modelling the Cosmic Star-Formation Rate}

Results from the CFRS (Lilly et al.\ 1998) tell us that while large disks show increases in
SFR by factors of $\sim$3 by z$=$0.7, the red (spheroid) components show no evolution. We
shall interpret the above observations as `spheroids form early and disks form late' (i.e. we consider the hump at $\sim$10 Gyrs in Fig.\ 1 as predominately
due to evolution of the spheroid component while the plateau at shorter
lookback times to be due predominately to disk evolution).
 
In order to model
the spheroid component we shall use the one-zone mass-loss chemical evolution
model of Hartwick (1976) which has been shown to reproduce the metal abundance
distribution of Galactic halo components quite well (e.g.\ Fig.\ 1 of Hartwick 1983). The 
model predicts a metal abundance (Z) distribution given by
\vskip 18pt
\centerline{dS/dlogZ = ln10$\times$S$_{tot}$$\times$(Z/P$_{eff}$)$\times$e$^{-Z/P_
{eff}}$~~~~~~~~~~~~~(1)}
\vskip 18pt
\noindent
where P$_{eff}$ is the effective yield and varies inversely as the star
formation (mass loss) rate, S is the cumulative number of stars formed and S$_{tot}=$ S$_{Z=\infty}$.

For a given age-metallicity relation (AMR), 
we can calculate the star-formation rate by taking the product of eqn 1 with
the derivative of the AMR, dlogZ/d$t$. 
 An AMR was constructed by assuming an age of 
13 Gyrs for M92 ([Fe/H]$=-$2.3), 11 Gyrs for M5 ([Fe/H]$=-$1.4), and
passing through [Fe/H]$=$0 at $t=0$. The resulting expression (shown dashed in Fig. 5) is
\vskip 18pt
\centerline{log Z/Z$_\odot$$=$0.117531$\times$(1$-$e$^{t_{9}/4.3}$)~~~~~~~~~~~~~~~~~~~~~~~~~~~(2)}
\vskip 18pt
\noindent
where $t_{9}$ is the lookback time in Gyrs.

For an assumed value of P$_{eff}$$=-$1.0 and S$_{tot}$=3.9$\times$10$^{10}$ M$_\odot$,
the resulting SFR is shown in Fig. 2. Note that the maximum SFR of $\sim$10 
M$_\odot$~yr$^{-1}$ (which depends only on the value of S$_{tot}$ and the AMR slope) is in the range of values deduced by Steidel et al.\ (1996) for the $z > 3$ `UV-dropout' galaxies.

\begin{figure}
\plotfiddle{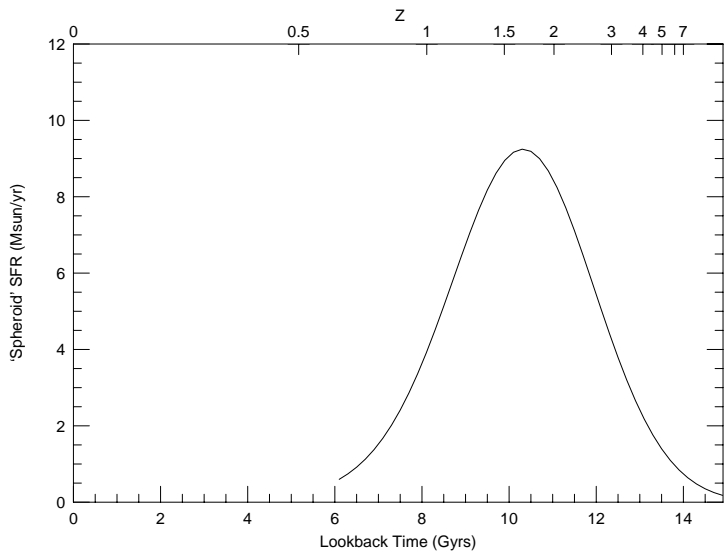}{2.5in}{90}{40}{40}{144}{-25}
\caption{The star-formation rate for the spheroid component of the model.}
\end{figure}

For the disk component we are guided by the SFR derived recently for the Galactic disk by Rocha-Pinto $\&$ Maciel (1997) which is shown in Fig. 3. For
ease of computation we have adopted the relation shown by the two straight 
line segments. (Note the increase in SFR by a factor of $\sim$3.4 between $t=0$ and $t=9$ Gyrs.)

\begin{figure}
\plotfiddle{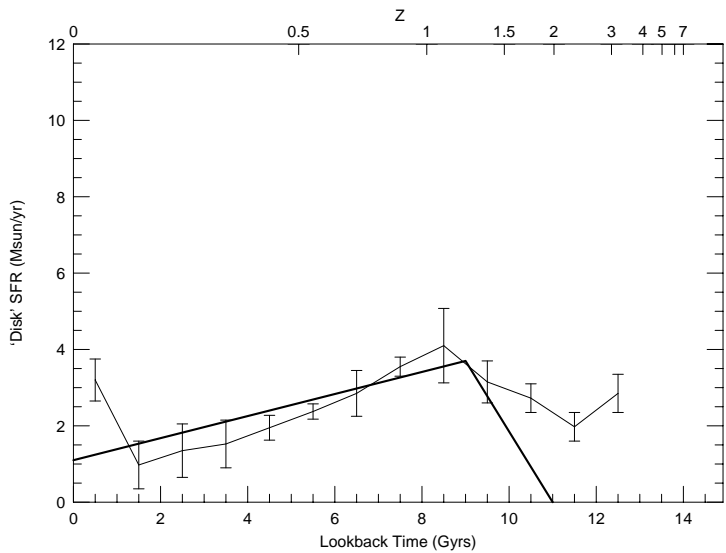}{2.5in}{90}{40}{40}{144}{-25}
\caption{The adopted star-formation rate for the disk component of the model (heavy lines) plotted on top of the derived rate for the Galactic disk from Rocha-Pinto \& Maciel (1997).}
\end{figure}

The disk SFR of Fig. 3 can now be added to the spheroid SFR component of
Fig. 2 and compared with the cosmic SFR of Fig. 1. This is shown in Fig. 4. 
Given the simplicity of the model, the agreement seems quite remarkable! 
We should emphasize that the composite galaxy modelled above is not expected to exist as a single object in nature. Rather, if the universe contained  only one galaxy that is what it might look like!

\section{Discussion and Predictions}

As indicated above, the star-formation rate of each component was chosen to
fit the relative SFR of Fig.\ 1. In doing so the model makes several predictions.
The value of S$_{tot}$ required to match the observed SFR implies a total baryon
mass of 18.5$\times$10$^{10}$ M$_\odot$ and a total mass of long-lived stars and
remnants of 3.1$\times$10$^{10}$ M$_\odot$. Integrating the assumed disk SFR of
Fig.\ 3 yields a stellar mass of 2.5$\times$10$^{10}$ M$_\odot$. Allowing an 
additional $10\%$ of the disk mass in gas we calculate the ratio of spheroid to disk mass of 1.1. Schechter $\&$ Dressler (1987) have determined this ratio   observationally and find it to be $\sim$1. Further, the ratio of total baryon
mass to mass in stars is 3.3. Equating this ratio to $\Omega_{baryon}$/$\Omega_{stars}$ and assuming $\Omega_{stars}$=0.003 (Persic $\&$ Salucci 1992) we find $\Omega_{baryon}$=0.01 compared to $\Omega_{BBN}$=0.03. Thus, the model predicts an alleviation of the `dark baryon problem' by putting 2.3 times the mass in stars into the IGM as metal-enriched
gas. Fig.\ 5 shows how the enrichment of the IGM is expected to vary as a function of lookback time. Note that the present metallicity of the IGM is
predicted to be 0.1~Z$_\odot$ (assuming that it is well mixed).

\begin{figure}
\plotfiddle{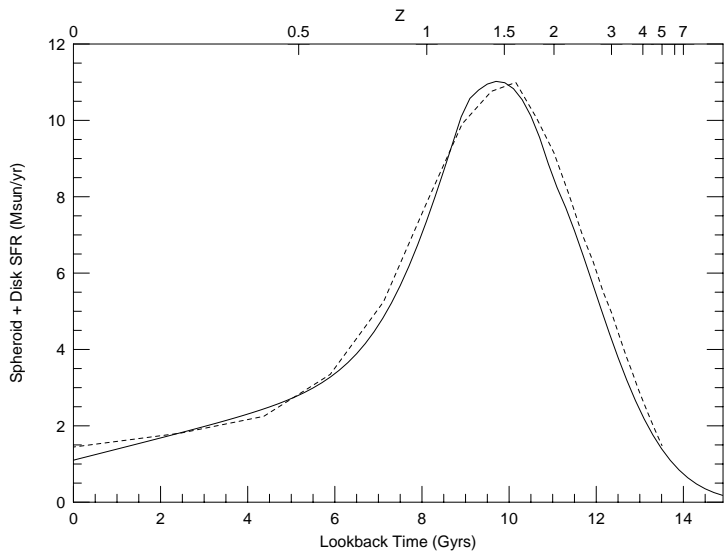}{2.5in}{90}{40}{40}{144}{-25}
\caption{The model spheroid component of Fig.\ 2 and the model disk component of Fig.\ 3 combined (solid line) compared to the observed relation (dashed) from Madau et al.\ (1998) in Fig.\ 1.}
\end{figure} 

How robust are these results? The cosmological model dependence is expected to be small, amounting basically to an expansion or
contraction of the abscissae in Figs.\ 1-5. However drastic changes to the original Madau et al.\ (1998) cosmic star-formation rate will have an appreciable effect. For example, if recent suggestions (Hughes et al.\ 1998) of a factor of 5 increase in SFR at redshift z$\sim$3 are confirmed, then the spheroid contribution would have to rise substantially and increase the derived spheroid to disk ratio upsetting the already good agreement with available observations. However, if the dust-enshrouded star-formation component is more nearly evenly distributed in redshift, then our results should remain valid, requiring only a change in normalization.

\section{How Do the Milky Way's Satellites Fit into This Picture?}

The above model does not differentiate between star formation taking place in one large object or in many smaller ones. Indeed, star formation followed by mass loss may also be effective if it takes place during collisions between  large numbers of smaller objects. It is easily shown that if the collisions which suddenly heat and/or sweep out the gas are Poisson distributed about mean Z$=$P$_{eff}$, then the expected metallicity distribution is virtually identical to eqn (1). Furthermore, collisions are likely to initiate bursts of star formation as well as eventually to quench them. This picture appears to favor a hierarchical galaxy formation process (e.g.\ Searle \& Zinn 1978) whereby numerous small sub-units collapsed, began forming stars, then collided during assembly over an extended period of time. The gas removed by collisions between the building blocks would then either be lost to the IGM and/or settled down later to form the disk. The removal of gas might then be expected to leave an object resembling a dwarf spheroidal galaxy, if it could survive the sudden mass loss and potential tidal disruption.
The large age spread observed from deep color-magnitude diagrams of individual dwarf spheroidal galaxies (e.g.\ Da Costa 1997) could be taken as evidence for the prolonged spheroid assembly seen in Fig.\ 2. 

\begin{figure}
\plotfiddle{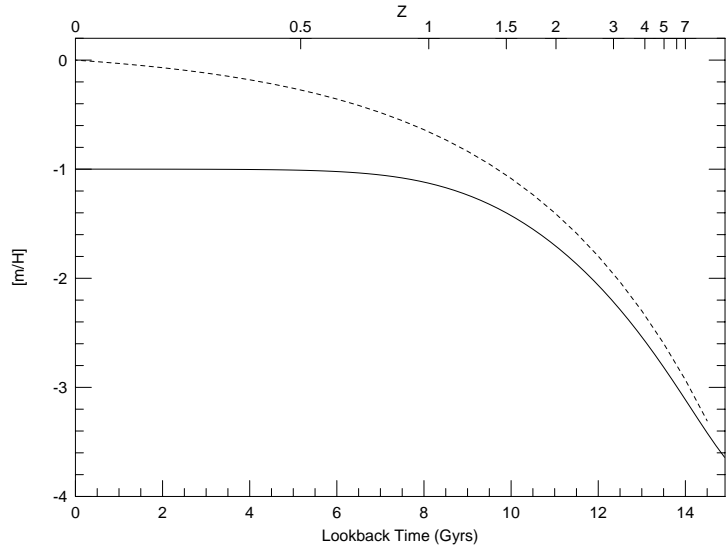}{2.25in}{90}{40}{40}{144}{-25}
\caption{The predicted metallicity of the gas lost during spheroid star formation (solid line). A fixed fraction of this gas forms the disk while the rest becomes the IGM. Dashed line: the assumed age-metallicity relation of eqn 2.}
\end{figure}

The Large Magellanic Cloud would appear to be a candidate for a massive leftover building block given that its most metal-poor clusters appear to be identical to those of the Galaxy (Olsen et al.\ 1998; Johnson et al.\ 1998). Such a picture would also allow a rationalization of the disk-like motions of the LMC clusters (Schommer et al.\ 1992) with the pressure-supported kinematics of the most metal-poor Galactic globular clusters.
 
Apparently the Magellanic Clouds have so far managed to avoid major mass loss by collision/star formation so that in the above scenario they would be considered  relics, having remained relatively self-contained from the time of formation of the first stars. The other gas-rich dwarf irregular galaxies in the Local Group may be similar examples.

\acknowledgements

The author is most pleased to have participated in this celebration of Sidney van den Bergh's monumental contributions to astronomy. He also wishes to thank colleagues S. Gwyn, M. Hudson, J. Navarro, and D. Vandenberg for many useful discussions. Finally the author gratefully acknowledges continued support from NSERC. 

\end{document}